\documentstyle[aps,psfig]{revtex}

\begin{document}

\title{On the genealogy of a population of biparental individuals}

\author{Bernard Derrida$^*$, Susanna C. Manrubia$\dag$, and
Dami\'an H. Zanette$\ddag$}

\address{
$^*$ Laboratoire de Physique Statistique de l'\'Ecole Normale
Sup\'erieure\\ 24 rue Lhomond, F-75231 Paris 05 Cedex, France\\
${\dag}$ Fritz-Haber-Institut der Max-Planck-Gesellschaft\\
Faradayweg 4-6, 14195 Berlin, Germany\\
${\ddag}$ Consejo  Nacional de  Investigaciones  Cient\'{\i}ficas  y
T\'ecnicas\\ Centro At\'omico Bariloche  e  Instituto  Balseiro\\
8400  S.C.  de Bariloche, R\'{\i}o Negro, Argentina}

\date{\today}

\maketitle

\subsection*{SUMMARY}
If one goes backward in time, the number of ancestors of an individual
doubles at each generation. This exponential growth very quickly exceeds the 
population size, when this size is finite. As a consequence, the ancestors 
of a given individual cannot be all different and most remote ancestors are 
repeated many times in any genealogical tree.
The statistical properties of these repetitions in genealogical 
trees of individuals for a panmictic closed population of constant size 
$N$ can be calculated. We show 
that the distribution of  the repetitions of ancestors 
reaches a stationary shape after a small number $G_c \propto \log N$ of
generations in the past, that only about $80\%$ of the ancestral population
belongs to the tree (due to coalescence of branches), and that two trees 
for individuals in the same population become identical after $G_c$
generations have elapsed.
Our analysis is easy to  extend to the case of exponentially growing population.

\section{Introduction}

In the case of sexual reproduction, the ancestry of an individual is formed 
by 2 parents, 4 grandparents two generations ago, and in general $2^G$
individuals $G$ generations back into the past. The explosive growth of
the number of ancestors belonging to the genealogical tree of, say, a present 
human should stop at some point due, at least, to the finite size of previous 
populations. For instance, only $G \simeq 33$ generations ago (spanning a 
period of less than thousand years), the number of potential ancestors in the 
tree of any of us is about $8.5 \times 10^9$, more than the present
population of the Earth, and of course much larger than the population living
about year 1000. The answer to this  apparent paradox is simple: The branches 
of a typical genealogical tree often coalesce, indicating that many of the
ancestors were in fact relatives and appear repeatedly in the tree  
(Ohno, 1996; Derrida {\it et al.}, 1999; Gouyon, 1999). 
It might be difficult to test the statistical properties of such
repetitions for an actual large, randomly mating population. Nevertheless, 
some exceptions can be found in royal genealogy.
Since nobles usually married within their own castes, the presence of
repeated ancestors in royal genealogical trees is far from rare. The
example of the English king Edward III, where some ancestors appear up to 
six times, has been analysed in our previous work (Derrida {\it et al.}, 
1999).\footnote{We used the tree of Edward III  which
can be found at http://uts.cc.utexas.edu/$\sim$churchh/edw3chrt.html.}

Much attention has been paid in the past to a related problem, namely the 
statistical properties of branching processes (Harris, 1963) and its 
applications to the characteristics of the successive descendants of a single
ancestor (Kingman, 1993). Actually, first applications of the branchig 
processes technique go back to the twenties.
J.B.S. Haldane (Haldane, 1927) calculated the probability that a mutant allele
be fixed in a population through a method developed previously by R.A. Fisher
(Fisher, 1922). There, the relevant quantity was the survival probability
of the descendants of the first individual carrying the mutation. All these 
studies apply to the vertical 
transmission of names, to the inheritance of characters coming only from 
one of the parents, like mithochondrial DNA or the Y chromosome, or to the fate
of a mutant gene, for example, and correspond to an effective monoparental 
population. The heart of our problem is to take into consideration that 
reproduction is biparental. The distribution of repetitions of ancestors 
described below does however satisfy an equation similar to those which appear
in branching processes (Harris, 1963).

Our problem of repetitions of ancestors in genealogical trees is much closer 
to the counting of the descendants of an individual in a sexual
population. For example, in the case of a population of constant size,
the average number of offspring is {\it two} per couple. Therefore
after $G$ generations each individual has on average $2^G$ descendants.
What prevents the number of descendants from growing exponentially with $G$ and
to exceed the population size is interbreeding: When $2^G$ becomes comparable 
to the population size, interbreeding happens between the descendants and 
different lines of descent coalesce. The problem of  the statistical properties 
of these coalescences is very similar  to our  present study   of genealogical 
trees. None of them has --to our knowledge-- yet been analysed.

In the present work, we study   theoretically the problem of repetitions in
the genealogical trees in the case of a closed, panmictic population. The 
study of the properties of a single 
tree with coalescent branches and the comparison of the genealogical trees 
of two contemporary individuals allows us to show that 

\begin{enumerate}
\item There is a finite fraction (about 20\% for a population of constant size 
$N$) of the initial population whose descendants becomes extinct after a number
of generations $G_c \propto  \log N$.
All the rest of the initial population (about 80\%) belongs to all genealogical 
trees, 
\item The distribution  of  the repetitions of ancestors living more than $G$ 
generations ago reaches a stationary shape after about $G_c $ generations,
\item The genealogical trees of two individuals in the same population become
identical after a small number of generations $G_c$ back into the past,
\item The similarity between two genealogical trees changes from 1\% (almost 
all ancestors in the two trees are different) to 99\% (the repetitions of the 
ancestors in the two trees are almost identical) within 14 generations 
around $G_c$, independently of the population size $N$.
\end{enumerate}

Our work can be generalized (see section IV) to describe coalescent processes,
understood as the study of the gene tree originated when looking for the 
ancestry of a random sample of sequences (Kingman, 1982; Hudson, 1991; 
Donnelly \& Tavar\'e, 1995). In the absence of
recombination, each sequence has a single ancestor. The topology of thus
reconstructed trees is equivalent to that generated through branching 
processes.
Next in complexity, one can consider a two-loci sequence and assume that 
recombination can occur only between the two loci and with a small 
probability (meaning correlated genealogies\footnote{In this paper, we use 
the term {\it genealogy} to refer to the
ancestry of a single gene or of a whole set of sequences. In all cases,
the {\it genealogy} is the complete set of ancestors contributing to the
present object, this  object  being an individual (as in section II),
a group of individuals (as in section III), a sequence (section IV), or a 
single locus (as quoted here). In this case, {\it correlated genealogies} 
simply means that the different sets of ancestors for the two-loci are 
not independent.}
for the two loci). The 
statistical properties of such process can be estimated until the most 
recent common ancestor (MRCA) is reached (Hudson, 1991). Instead, if 
one faces the study of a chromosome (Wiuf \& Hein, 1997; Derrida \& 
Jung-Muller, 1999) or of the whole genome,
the number of ancestors grows as one proceeds back in time, since each
individual has two parents and, apart from coalescence, also
recombination (meaning splitting of the branches in the tree) is frequent.

If one considers a population or a sample  of individuals within a 
population, there are relevant differences between the genealogy of a 
single gene and the genealogy of a chromosome or of the whole genome 
(which we study here). While in the first case, in fact, there exists
a MRCA for the sample (where the gene tree ends), the genealogical 
tree of a chromosome or of the genome with two parents proceeds 
backwards in time and never reduces to a single ancestor. The 
genealogical tree representing the pedigree of a diploid organism 
contains a large fraction of the ancestral population. In this case, 
one may then talk about the {\it most recent
common set of ancestors}, and study the similarities among different 
individuals now within the same population.

\section{Statistical properties of an individual tree}

Here
we consider a simple neutral model of a closed
population evolving  under  sexual  reproduction and with  non-overlapping 
generations.\footnote{The Wright-Fisher model for allele frequencies 
works in the same
set of hypothesis (Wright, 1931; Fisher, 1930). More recently, Serva and
Peliti (Serva \& Peliti, 1991) obtained a number of statistical results
for the genetic distance between individuals in a
sexual population evolving in the absence of natural selection.}
If the population size is $N(G)$ at generation $G$ in 
the past, we form couples at random (by randomly choosing $N(G)/2$ pairs 
of individuals) and assign each couple a random number 
$k$ of descendants. The probability $p_k$ of the number $k$ of offspring is 
given and if the population size is $N$ at present, its size $N(G)$ 
at generation $G$ in the past is given by
\begin{equation}\label{NG}
N(G)= \left({2 \over m}\right)^G N 
\end{equation}
where the factor $m$ is obtained from
\begin{equation}
m= \sum_k \ k \ p_k \; .
\end{equation}

For $m=2$, the  population size 
remains constant in time, whereas for $m \neq 2$ the number of individuals 
in the next generation is multiplied by a factor $m/2$. After a number 
of generations, the tree of each of the individuals in the 
youngest generation is reconstructed. To quantify the contribution of each 
of the ancestors to the genealogical tree of an individual, we define the 
{\it weight} $w_\gamma^{(\alpha)}(G)$ of an ancestor $\gamma$ in the tree of 
individual $\alpha$ at generation $G$ in the past as 

\begin{equation}\label{weight}
w_{\gamma}^{(\alpha)}(G+1)= {1 \over 2} \sum_{\gamma' \;\;\; \makebox{children 
of} \;\;\;  \gamma} w_{\gamma'}^{(\alpha)}(G)
\label{recursion}
\end{equation}
We take $w_{\gamma}^{(\alpha)}(0)= \delta_{\alpha,\gamma}$,
as this ensures that  at generation $G=0$ all the weight is carried by the 
individual $\alpha$ itself. 
The factor $1/2$ in (\ref{recursion}) keeps the sum of the weights normalized
$\sum_{\gamma=1}^{N(G)} w_{\gamma}^{(\alpha)}(G)=1$, for any past generation 
$G$. The weight  $w_\gamma^{(\alpha)} (G)$ can 
be thought of as the probability of reaching ancestor $\gamma$ if one climbs 
up the reconstructed genealogical tree of individual $\alpha$  by choosing at 
each generation one of the two parents at random. The weights essentially 
measure the repetitions (see figure 1) in the genealogical tree. Without 
repetitions, $w_\gamma^{(\alpha)}(G)$ would  simply  be $2^{-G}$ for each 
ancestor $\gamma$ in the tree.

As an illustration of the previous quantities, we represent in Fig. 1
the result of random matings inside a small closed population of constant 
size $N=14$ (thus $m=2$) during 7 generations. The lines link progenitors
with their offspring. The grey scale gives the weight $w_{\gamma} (G)$ of
each of the individuals in the tree. The
numbers on the left, all of them of the form $r/2^G$, give the weight of the 
leftmost individual in each generation. The denominators simply indicate the 
potential maximum number of ancestors at each generation. As counted by the 
numerator, each of them would appear repeated $r$ times in this tree if all 
the branches were explicitly shown. 

We further assume that the probability $p_k$ of having $k$ children per
couple
follows a Poisson distribution, $p_k=m^k e^{-m}/k!$ (most of what follows 
could be easily extended to other choices of $p_k$). We represent in Fig. 
2 the probability for 
an English couple to have $k$ marrying sons during the period 1350-1986 
(Dewdney, 1986). The solid line corresponds to a Poisson distribution with 
average
$1.15$ (i.e., the average number of offspring {\it per individual} in that 
period, which corresponds to $m=2.3$ in our analysis), and implies that the 
total population is growing. These data spanning six centuries and taken over 
an homogeneous population support the hypothesis that the number of offspring 
is indeed Poisson distributed.\footnote{Nonetheless, deviations from this 
distribution induced by a social 
transmission of the reproductive behaviour have been reported (Austerlitz
\& Heyer, 1998).}

If we define $S^{(\alpha)} (G)$, the fraction of the population (at a generation
$G$ in the past) which does not belong to the genealogical tree of individual 
$\alpha$, (i.e. such that $w_{\gamma}^{(\alpha)}(G)=0$) one can show
(see the appendix) that

\begin{equation}\label{SG}
S^{(\alpha)} (G+1) = \exp \left[ -m + m \  S^{(\alpha)} (G) \right] \; .
\end{equation}
This  recursion, together with the initial condition $S^{(\alpha)} (0)=1 - 
1/N$, determines this quantity for any $G$ (Derrida {\it et al.}, 1999).

For large $G$ and for any individual $\alpha$, this fraction 
$S^{(\alpha)} (G)$ converges to the fixed point 
$S(\infty)$  of (\ref{SG}). This gives for $m=2$ (i.e. for a population of 
constant size) a fraction $S(\infty) \simeq 0.2031878..$ 
which becomes extinct, 
so that the remaining  fraction $1-S(\infty) \simeq 80 \%$ of the population 
belongs to the genealogical tree {\it of any individual} $\alpha$. A similar 
calculation shows that this 80\% of the population which is not extinct 
after a large number of generations appears in the genealogical trees of 
all individuals: If $S^{(\alpha,\beta)}(G)$ is the fraction of the 
population which does not belong to any of the two trees of two 
distinct individuals $\alpha$ and $\beta$,  $S^{(\alpha,\beta)}(G)$ 
satisfies the same recursion (\ref{SG}) as $S^{(\alpha)} (G)$, and 
converges to the same fixed value $S(\infty)$. Thus, within this neutral
model, an individual either becomes 
extinct (with a probability of 20\%) or becomes an ancestor of the whole 
population after a large number of generations (with a probability of 80\%).
For an exponentially growing population with $m=2.3$ as in figure 2, the 
results are the same except for the precise value 
of $S(\infty)$ (for $m=2.3$, one finds $S(\infty) \simeq 14\%$).

When $G$ is large enough, as shown in the appendix, the whole distribution 
$P(w)$ of the weights $w_\gamma^{(\alpha)}(G)$  reaches a stationary shape,
the properties of which
can be calculated (Derrida {\it et al.}, 1999). We show in Fig. 3 the 
distribution $P(w/ \langle w \rangle)$ for 
different values of $m$. As can be seen, it has a power-law dependence, $P(w) 
\propto w^{\xi}$ for small values of the ratio $w/\langle w \rangle$, 
with an exponent given by 
\begin{equation}
\xi=- {\log S(\infty) \over \log m} -2 \; ,
\label{beta}
\end{equation}
and achieves a maximum value for $w/{\langle w \rangle} \simeq 1$.

\section{Similarity between two trees}

We would like to know how similar are the genealogical trees of two 
contemporary individuals and how they evolve in time within the same
population. We have seen that a large fraction $1-S(\infty) \simeq 80\%$ of 
the ancestral population constitutes the pedigree of every present
individual. As a next step, one can compare two individuals and compute
the degree of similarity between their trees, that is, the set of
ancestors appearing at each generation in both trees simultaneously. 
We will see in particular that the two trees become identical after a 
number $G_c$ of generations. 

We start with the definition of the 
{\it overlap} between  the genealogical trees of two different individuals,
$\alpha$ and $\beta$. Let $w_{\gamma}^{(\alpha)}(G)$ be the weight of the 
ancestor $\gamma$ in the tree of $\alpha$ at generation $G$ in the past, and 
similarly let $w_{\gamma}^{(\beta)}(G)$ be the weight of the same ancestor 
$\gamma$ at generation $G$ for $\beta$. These weights evolve according to 
(\ref{weight}) with $w_{\gamma}^{(\alpha)}(0)=\delta_{\gamma,\alpha}$ and 
$w_{\gamma}^{(\beta)}(0)=\delta_{\gamma,\beta}$ at generation $G=0$.
In order to quantify the similarity between the two trees, we introduce the 
quantities

$$X^{(\alpha)}(G)=
\sum_{\gamma=1}^{N(G)}  \left[ w_{\gamma}^{(\alpha)}(G) \right]^2 $$
and
$$Y^{(\alpha,\beta)}(G)= \sum_{\gamma=1}^{N(G)}  w_{\gamma}^{(\alpha)}(G)  \ 
w_{\gamma}^{(\beta)}(G) \ .  $$
$Y^{(\alpha,\beta)}(G)$ measures the  correlation between the two trees  at 
generation $G$ in the past and $X^{(\alpha)}(G)$ acts as a normalization 
factor.  We then define the overlap $q^{(\alpha,\beta)}(G)$ between the two 
trees at that generation  by

$$q^{(\alpha,\beta)}(G) = {Y^{(\alpha,\beta)}(G) \over  \left[ X^{(\alpha)}(G)
\ X^{(\beta)}(G) \right]^{1/2}} $$
This overlap is a measure of the (cosine of the) angle  between the two 
$N-$dimensional vectors $ w_{\gamma}^{(\alpha)}(G)$ and $w_{\gamma}^{(\beta)}
(G)$.\footnote{Similar quantities have been proposed
as an indicator of the amount of evolutionary divergence between populations 
(Kimura, 1983). The quantity  analogous to our weight 
$w_{\gamma}^{(\alpha)}$ in the population genetics approach is the
frequency of the sampled alleles, the number of ancestors $\gamma$ corresponds to
the number of genes (that is the dimension of the space in which the
vector $w_{\gamma}^{(\alpha)}$ is embedded), and our individuals $\alpha$
and $\beta$ correspond to the compared populations (Cavalli-Sforza \& 
Conterio, 1960).}
When $q^{(\alpha,\beta)}(G) \simeq 0$, the two vectors are essentially 
orthogonal and  the ancestors of $\alpha$ and $\beta$ are all different. On the 
other hand, when $q^{(\alpha,\beta)}(G) \simeq 1$, the vectors are almost 
identical (as for brothers).

For a large enough population, the fluctuations of $X^{(\alpha)}(G)$ and 
$Y^{(\alpha,\beta)}(G)$ are small around the population averaged values 
$\langle X(G)\rangle $ and $\langle Y(G) \rangle$ for almost all choices 
of $\alpha $ and $\beta$. Of course, 
if $\alpha$ and $\beta$ are brothers, $Y^{(\alpha,\beta)}(G)=X^{(\alpha)}(G)$,
a value very different from its average $\langle Y(G) \rangle $; it is however
very unlikely to get brothers, sisters or even cousins if one picks up  two 
individuals at random from a large population.

The averages $\langle X(G) \rangle$ and $\langle Y(G) \rangle$ can be 
calculated from the evolution of the weights (\ref{recursion}). 
Initially, $X(0)=1$ and $Y(0)=0 $ since the individuals  $\alpha$ and
$\beta$ in any pair are different.
Using the fact that for large $N$ the fluctuations of $X^{(\alpha)}(G)$ and 
$Y^{(\alpha,\beta)}(G)$ are 
small, the expected value of the overlap $q(G)$ between two randomly chosen 
individuals is given by

\begin{equation}\label{qab}
q(G)\  \simeq \ {\langle Y(G) \rangle \over \langle X(G) 
\rangle } \  = \  {1 \over 1 + m^{G_c -G}} $$
\end{equation}
where
\begin{equation}\label{ngen}
G_c = {\log ((m-1) N ) \over \log m} - 1 \; .
\end{equation}
This expression is derived in the appendix.
Of course Eq (\ref{qab}) is only valid with probability one with respect 
to the random choice of $\alpha$ and $\beta$ and with respect to the 
dynamics.
We see that for large $N$, the  overlap $q(G)$ is essentially zero for a number 
of generations of order $G_c \simeq \log N / \log m$ and then within a number 
of generations $\Delta G$ which {\it does not depend on $N$}, it becomes equal
to unity. Fig. 4 displays the averaged overlap $q(G)$  as a 
function of the number of generations $G$ for different values of $N$. We have
chosen $m=2$ so that the population remains constant in size. We see that 
changing $N$ does not change the $G$ dependence except for a translation 
of the curve. In particular the range  $\Delta G$ on which the overlap 
changes from $1 \%$ to $99 \%$ does not depend on $N$. It is easy to check 
from (\ref{qab}) that for $m=2$, the overlap should satisfy 
\begin{equation}
q(G+1)= { 2 
q(G) \over 1 + q(G)}
\label{qevo}
\end{equation}
(plain line in the insert). The fixed point $q(G)=0$ is 
unstable for this map. All the trajectories finally converge to the stable 
fixed point $q(G)=1$ for large $G$. Also the quantity 
$\Delta G$ can be estimated by counting how many generations are required 
for the overlap to  change from 1 \% to 99 \% and this gives from (\ref{qab})

$$\Delta G \simeq  \log (10^4) /\log m,$$ 
that is $\Delta G \simeq 14$ for $m=2$ and $ \Delta G \simeq 11$ for $m=2.3$ 
as in figure 2. Typical values of $G_c$ are $G_c \simeq 20$ for a population 
of constant size $N=10^6$. For a population increasing  with $m=2.3$ as in 
figure 2, one gets $G_c=21 $ if the size  in the last generation is $N=N(0)=75$ 
millions.

The previous analysis can be easily extended to the hypothetical case of 
having an arbitrary number $p$ of parents instead of 2. As is shown in 
the appendix, the statistical properties of genealogical 
trees in a population of constant size but arbitrary $p$ are the same as 
for a population with only two parents and an expanding or shrinking
size according to Eq. (\ref{NG}). The described 
statistical properties are thus equivalent in (i) a system with 
sexual reproduction and a growth rate $m=p$ and (ii) a system with 
constant population size but a number $m$ of genders.

The existence of a  generation  $G_c$ around which the genealogical similarity 
among individuals changes from 0 to 1 and which grows logarithmically with the 
size of the population is one of our main results. This has to be compared with 
the number of generations required  for the population to become genetically  
homogeneous (Donnelly \& Tavar\'e, 1991; Harpending {\it et al.}, 1998), 
which grows proportionally to $N$.
The difference is that when $G_c  \ll G \ll N$, all the overlaps are 1, i.e
all the genealogical trees in the population have the  same ancestors with the 
same weights, but the genomes are still very different: This is just an 
extension of the situation of  brothers who have exactly the same genealogical
tree but different genomes. 

\section{Simple model for the contribution of the ancestors to the genome}
The evolution of a set of sequences subject to coalescence and
recombination was first described by Hudson (1983). In this case,
evolution proceeds until the most recent common ancestor for each
set of homologous sites has been found. The set of MRCA sites does not
necessarily belong to the genome of a single ancestor, on the contrary,
it is in general spread on a finite fraction of the original population
(Wiuf \& Hein, 1997; 1999). In this section, we focus our attention on
the statistical properties of the ancestry of a single extant genome. In
particular, we calculate the equilibrium distribution for the fraction
of material contributed by each ancestor. 

Consider the whole set of genes 
that a present diploid organism has inherited from its parents. Although 
both parents contributed $50\%$ each, it is no longer true that 
grandparents contributed $25\%$ each, since independent assortment of 
chromosomes plus crossing
over mixed in each of the parental gametes the material inherited from
the previous generation. 
As a rough approximation to the output of genetic recombination, one 
might consider that each sequence is obtained as the addition of a 
fraction $f$ of the genetic material of one parent and a fraction 
$1-f$ of the genetic material of the other parent with  $f \in (0,1)$. 
This would be true if the length $L$ of the sequence was long enough
(or infinitely long), so that there would be no restriction on the 
number of times it could be divided, and if one could forget the linear
structure of the sequence.
The process of coalescence and recombination (for small $N$) is 
schematically represented in Fig. 5. 

We can now repeat the analysis done previously to the present extension.
We will discard the correlations between the values of $f$ coming from
a couple. This is equivalent to our assumption that fixing the pairs for
$k$ offspring or choosing the parents of each individual at random only
has effects of order $O(N^{-1})$ (see the appendix), and we can therefore
work in the simplest realization of the process. Hence, we assume that
the fraction $f$ takes independent values for each parent. The 
recursive equations (\ref{weight}) for the weights become

\begin{equation}
w_{\gamma}^{(\alpha)} (G+1) = \sum_{\gamma' \; \text{children of} \;
\gamma}
f_{\gamma'} \ w_{\gamma'}^{(\alpha)}(G) \; ,
\label{wevo}
\end{equation}
where the weight $w_{\gamma}^{(\alpha)} (G)$ means now the fraction of 
the genetic material of individual $\alpha$ inherited from ancestor 
$\gamma$ at generation $G$.
The random fraction $f$ is chosen anew for each offspring from
a distribution $\rho(f)$ (with average value $\langle f \rangle = 1/2$). 
This implies that now even brothers would have different weights for their 
ancestors, and hence brings us slightly closer to the real genetic process. 

Following the procedure described in the appendix, one
can calculate the fraction $S$ of ancestors without lines of descent in
the present (as we also show in Sec. II)
and the exponent $\xi$ for the distribution $P(w)$. In
general, given the distribution $\rho(f)$ for the contributions of the
parents, we get
\begin{equation} \label{Sm}
S(\infty)  = e^{m S(\infty) -m}
\end{equation}
\begin{equation}\label{rho}
1= S(\infty) \   m^{2 + \xi}  \  \langle f \rangle^{1 + \xi}  \   \int f^{-\xi-1} \rho(f) \text{d}f \; .
\end{equation}
as one can easily show from  (\ref{wevo}) that the generating function 
$h_G(\lambda)$ defined by $h_G(\lambda) = \langle \exp[ \lambda w(G) / 
\langle w(G) \rangle ] \rangle $ has a limit  $h_\infty(\lambda)$ for 
large $G$ which satisfies

$$h_\infty(\lambda) = \exp \left[ -m + m \int \rho(f) \text{d} f \ h \left( 
{\lambda f \over m \langle f \rangle } \right) \right] $$

Fig. 6 summarizes the changes in the distribution $P(w)$ for different
distributions $\rho(f)$ of the random variable $f$. We have
considered a simple case of a population  of constant size (i.e. $m=2$) 
and  with $\rho(f)=1/(2\delta)$ uniform 
in  the interval $(1/2-\delta,1/2+\delta)$. In this particular
case, an implicit relation between $\delta$ and the exponent $\xi$ can be
obtained,

\begin{equation}\label{bd}
\delta \xi = S \left[ \left({1 \over 2} - \delta \right)^{-\xi} -
\left({1 \over 2} + \delta \right)^{-\xi} \right] \; .
\end{equation}
As $\delta $ varies, $P(w)$ remains a power law at small $w$ (i.e. $P(w) 
\propto w^{\xi}$), and
the exponent $\xi$ monotonously decreases with $\delta$. In
particular, for $\delta \simeq 0.35$, $\xi$ changes sign: The maximum of 
$P(w)$ moves discontinuously from $w/{\langle w \rangle} \simeq 1$ to 
$w/\langle w \rangle \simeq 0$. 
The exponents obtained through simulations of the process are represented 
in Fig. 7 together with the numerical solution of Eq. (\ref{bd}), showing 
a good agreement.

\section{Discussion}

We have analysed the statistical properties of genealogical trees
generated inside a closed sexual population. We focused our  interest  
on the distribution of the repetitions of ancestors in 
the trees and on the amount of genetic material contributing to
an extant genome. 
The precise values of $\xi, S(\infty), G_c$ and $\Delta G$ depend 
only weakly on the details of the model and do not change qualitatively if 
for instance a non Poissonian distribution of offspring is used.
Moreover, we have shown how our results can be extended to the
hypothetical case of having an arbitrary number $p$ of parents: Indeed,
this case proves to be equivalent to a biparental population with a
growth rate $m/2=p/2$. 

The problem  analysed here presents a number of connections to other fields. 
Equations similar to (\ref{recursion}) appear also in the distribution of 
constraints in granular media where the variables $w$ represent the force 
acting on each grain and the recursion (\ref{recursion}) expresses the way in
which constraints are transmitted from one layer to the next (Coppersmith
{\it et al.}, 1996). 
In this case, $p \neq 2$ and even fluctuating $p$ would be perfectly realistic.
The fact that the overlap  changes from 0 to 1 within a small number of 
generations $\Delta G$ independent of the size of the population and after 
$G_c \simeq \log N$ generations is also very reminiscent of the sharp cutoff 
phenomenon characteristic of some natural mixing processes modelled by Markov 
chains. One example of such systems is the shuffling of cards, where the 
stationary state in which the system has lost almost all information about the 
initial ordering of the $n$ cards is reached through a sharp cutoff after 
about $\log n$ riffle shuffles (Diaconis, 1996). 

It is clear that the study of  the interplay between the weights calculated 
in our generalized model and the structure of the genome would require more 
sophisticated approaches (Derrida \& Jung-Muller, 1999; Wiuf \& Hein, 1997; 
1999). We have discarded the correlations between the history of neighboring 
sites in a sequence and assumed the independence of the factors $f$.
Actually, the closer in the sequence two positions are, the more
correlated are their genealogical histories (Kaplan \& Hudson, 1985).
This fact constrains the possible breaking points for our simulated 
sequences, implying that the random factors $f$ in (\ref{wevo}) are a crude 
approximation to reality. 

Since we have faced the problem from a statistical perspective, our 
results represent the average, typical behaviour, and are only valid with 
probability one when the population size is large. We did not study 
fluctuations due to the finite size of the population. Nonetheless, we 
hope that our results contribute to a better understanding of the role of 
genealogy in the degree of diversity of finite-size interbreeding populations.

\section*{Appendix}

In this appendix we have regrouped the  technical aspects of the derivations
of the main equations (\ref{SG},\ref{beta},\ref{qab},\ref{Sm},\ref{rho}) 
presented in the body of the paper.

One may consider several  variants of the model which all give a Poisson 
distribution for the number of offspring when the size of the population is 
large. For instance,
the population size could be strictly multiplied by a factor 
$m/2$ at each generation or it could fluctuate (if we take the number of 
offspring from the Poisson distribution). One might decide that each 
individual has two parents chosen at random in the previous generation or 
form  fixed couples and assign each couple some children.
All these variants do not change the results  when the population size
is large, but might affect some  finite size corrections that we compute in 
this appendix.

We will choose the following version of the model, which makes the calculation
of the finite size corrections not too difficult. Our population has a  
given size $N(G)$ at each generation $G$ in the past, and
we will assume that all the $N(G)$ are very large, at least in the range
of generations $G$ that we will consider.
Now, to construct the ancestors of all 
the $N(G)$ individuals at generation $G$ in the past, we choose  for each 
of them a 
pair of parents at random among the $N(G+1)$ individuals at the previous 
generation (to facilitate the calculation, we do not even exclude that the 
two parents might be equal).
Within this model,  the number  $k$ of children of an individual
at generation $G+1$ is random and can be written as
$$k= \sum_{i=1}^{2 N(G)}  z_i$$
where $z_i=1$ with probability $1 / N(G+1)$ and $z_i=0$ otherwise.
It follows that the whole distribution of $k$ can be calculated.
The probability  $p_k$ that an individual at  generation $(G+1)$ has exactly 
$k$ children is  given by the binomial distribution
\begin{equation}
p_k = { (2 N(G)) ! \over k ! \  ( 2 N(G) -k)!} \left( 1 \over N(G+1) \right)^{k}
\left( 1 - {1 \over N(G+1) }\right)^{ 2 N(G)-k}  \; .
\label{pkexact}
\end{equation}
In particular,
\begin{eqnarray}
&&\langle  k \rangle =  {2 N(G) \over N(G+1) }
\nonumber \\
&&\langle  k(k-1) \rangle =   {2 N(G) [2 N(G) -1] \over N(G+1)^2 }
\nonumber \\
&&\langle  k (k-1) (k-2) \rangle =    {2 N(G) [2 N(G) -1] [2 N(G) - 2]  
\over N(G+1)^3 } \; .
\label{momk}
\end{eqnarray}

If the population size is multiplied by a factor $m/2$ at each generation,
i.e. if  $N(G)= N(G+1) \ m /2 $ (as $G$ counts the number of generations in 
the past), one recovers  from (\ref{pkexact})
the Poisson distribution  $p_k = m^k e^{-m} / k! $ for large $N(G)$.

\subsection{Calculation of the density of individuals without long term 
descendants and derivation  of (\ref{SG})}
To establish (\ref{SG}), one simply needs to notice that for an individual to 
have no descendants after $G+1$ generations, all his children should have no
descendants after $G$ generations. Let $M(G)$ be the number of individuals 
with no descendants at generation $G$ in the past.
Given $M(G)$, one can write $M(G+1)$ as

$$M(G+1) = \sum_{\gamma=1}^{N(G+1)} y_\gamma$$
where $y_\gamma =1$ if all the children  of $\gamma$ are among the $M(G)$ 
and $y_\gamma=0$ otherwise. It can be shown that 
$$\langle y_\gamma \rangle=\left(1-{1\over N(G+1)}\right)^{2N(G)-2M(G)}$$ 
and 
$$\langle y_\gamma y_{\gamma'} \rangle=
\left(1 - {2 \over N(G+1)} \right)^{2 N(G) - 2 M(G)}$$ 
for $\gamma \neq \gamma'$. This gives
\begin{equation}
\langle M(G+1) \rangle =N(G+1) \left( 1 - {1 \over N(G+1) }\right)^{2 N(G) 
- 2 M(G)}  
\label{M1}
\end{equation}
\begin{equation}
\langle M^2(G+1) \rangle = \langle M(G+1) \rangle + N(G+1) 
[N(G+1)-1]  \left( 1-{2 \over N(G+1)} \right)^{2N(G)-2M(G)}
\label{M2}
\end{equation}
When all the $M$'s and $N$'s are large, we see from (\ref{M1},\ref{M2}) 
that the fluctuations of $M(G+1)$ are small 
(as $ \langle M^2(G+1) \rangle - \langle M(G+1) \rangle^2 \ll
\langle M(G+1) \rangle^2$), and one finds from (\ref{M1})  that the ratio 
$M(G)/N(G) \equiv S^{(\alpha)}(G)$ satisfies 

$$ S^{(\alpha)}(G+1)=  \exp \left[ {2 N(G) \over N(G+1)} ( S^{(\alpha)}(G) -1 ) 
\right] $$
which is identical to (\ref{SG}) for $N(G) =  N(G+1) m/2 $.

\subsection{Time evolution of the  distribution of the weights }
From the recursion (\ref{weight}) and from the known distribution 
(\ref{pkexact}) of $k$ one can write recursions for the moments of the weights
\begin{eqnarray}
&&\langle  w_\gamma^{(\alpha)} (G+1) \rangle = 
 { \langle k \rangle \over 2}  \langle  w_\gamma^{(\alpha)} (G) \rangle
\label{w1} \\
&&
\langle [ w_\gamma^{(\alpha)} (G+1) ]^2 \rangle =
 {\langle k \rangle \over 4}  \langle [ w_\gamma^{(\alpha) } (G) ]^2 \rangle +
{\langle k(k-1) \rangle \over 4}
\langle w_\gamma^{(\alpha)} (G) w_{\gamma'}^{(\alpha)} (G) \rangle \; ,
\label{w2} 
\end{eqnarray}
where $\gamma \neq \gamma'$.
The normalization $\sum_\gamma w_\gamma^{(\alpha)} = 1$  allows one to rewrite

$$\langle w_\gamma^{(\alpha)} (G) w_{\gamma'}^{(\alpha)} (G) \rangle = {1 
\over N(G)-1}  \left[  \langle w_\gamma^{(\alpha)} (G)  \rangle -  
\langle [ w_\gamma^{(\alpha) } (G) ]^2 \rangle \right]  $$
and together with the known moments (\ref{momk}) gives that
\begin{eqnarray}
&& \langle  w_\gamma^{(\alpha)} (G+1) \rangle = 
{ N(G) \over N(G+1)}   \langle  w_\gamma^{(\alpha)} (G) \rangle = {1 \over 
N(G+1)}
\label{w1a} \\
&&
\langle [ w_\gamma^{(\alpha)} (G+1) ]^2 \rangle =
\left[ {N(G)  \over 2 N(G+1)}  - { N(G) \  [2  N(G) -1] \over 2 \   N(G+1)^2 \  
[N(G)-1]} \right] \langle [ w_\gamma^{(\alpha) } (G) ]^2 \rangle \nonumber \\
&& \;\;\;\;\;\;\;\;\;\;\;\;\;\;\;\;\;\;\;\; + {2 N(G) -1 \over 2 \  N(G+1)^2 \  [N(G)-1 ] }
\label{w2a} 
\end{eqnarray}
where $\gamma \neq \gamma'$.

For large $N(G)$, if the ratio $N(G+1)/ N(G) = 2 /m$, as in the case of a
population
increasing by a factor $m/2$ at each new generation, expression (\ref{w2a}) 
becomes simpler and one gets
\begin{eqnarray}
\langle [ w_\gamma^{(\alpha)} (G+1) ]^2 \rangle =
{m  \over 4}   \langle [ w_\gamma^{(\alpha) } (G) ]^2 \rangle +
{m^2 \over 4} \left( 1 \over N(G) \right)^2
\label{w2b} 
\end{eqnarray}
In this limit,  we have from (\ref{momk}) that $\langle k \rangle= m$ and
$\langle k (k-1) \rangle= m^2$, and we see that (\ref{w2b})  means that in 
(\ref{w1})   the weights $w_\gamma^{(\alpha)}$ and $w_{\gamma'}^{(\alpha)} $ 
are, for large $N(G)$,  uncorrelated. 
The calculation of higher moments of the weights can be done in the same manner
and  for large $N(G)$  the weights of different ancestors become again 
uncorrelated. 

If the population size changes in time, the distribution of the weights cannot 
be stationary. 
This is already visible in the expression (\ref{w1}) which shows that even
the first moment of the weights changes with $G$. 
One can however check from (\ref{w1}) and (\ref{w2b}) that the ratio 
$\langle [ w_\gamma^{(\alpha)} (G) ]^2 \rangle / \langle  w_\gamma^{(\alpha)} 
(G)  \rangle^2 $  which satisfies
\begin{eqnarray}
{\langle [ w_\gamma^{(\alpha)} (G+1) ]^2 \rangle 
\over \langle  w_\gamma^{(\alpha)} (G+1)  \rangle^2} =
{1 \over m} \    { \langle [ w_\gamma^{(\alpha) } (G) ]^2 \rangle
\over \langle  w_\gamma^{(\alpha)} (G)  \rangle^2  }
+ 1
\label{w2c} 
\end{eqnarray}
has a limit $m/(m-1)$  as $G$ increases. Moreover, as the initial value of this 
ratio is $N(0)$, the number of generations $G_c$  to converge to this limit is
$  G_c \sim \log N(0) / \log m$.
Higher moments of the weights behave in a similar way and one can write
recursions for ratios which generalize (\ref{w2c}) and which show that
all the ratios have limits.

This indicates  that the distribution of the ratio $w / \langle w \rangle$ 
becomes stationary. In the limit of large $N(G)$ (considering that the weights 
of the different children $\gamma'$ in (\ref{weight}) can be taken as 
independent and that the distribution of $k$ becomes Poissonian), one finds 
that the generating function $h_G(\lambda)$ defined by
\begin{equation}
h_G(\lambda) = \left\langle \exp \left[\lambda {w_\gamma^{(\alpha)}(G)  \over 
\langle w(G) \rangle} \right] \right\rangle 
\label{hl}
\end{equation}
satisfies
\begin{equation}
h_{G+1}(\lambda) =  \sum_{k} p_k \left[ h_G \left({\lambda  \langle w(G) 
\rangle \over 2 \langle w(G+1) \rangle } \right) \right]^k = \exp \left[  
- m + m \  h_G(\lambda /m) \right] \; .
\label{hl1}
\end{equation}
Recursion (\ref{hl1})
generalizes to the case $m \neq 2$ (i.e. the case of an exponentially 
increasing population) the result of our previous work obtained for a 
population of constant size ($m=2$). Similar recursions  have been studied 
in the theory of branching processes (Harris, 1963). 
The use of generating functions in population genetics is well illustrated in 
the book by Gale (1990), where this method is for example applied to the
calculation of the probability of fixation of a mutant allele.

It is remarkable, that if  one considers an imaginary world where each 
individual would have $p$ parents (instead of $2$), the generating function
(\ref{hl}), in the case of a population of constant size, would satisfy the 
recursion (\ref{hl1}) with $m=p$. This means that as long as the distribution 
of weights is concerned, the problem of a large population 
of constant size with $m$ parents per individual is identical to the problem 
of a population of  size increasing at each generation by a factor $m/2$ with
two parents per individual.

\subsection{Stationary distribution}
For large $G$, if we fix the ratio $N(G)/N(G+1)= m/2$, the generating function 
$h_G(\lambda)$ converges to $h_\infty(\lambda)$ solution of
\begin{equation}
h_\infty(\lambda) =   \exp \left[  - m + m \  h_\infty(\lambda /m) \right]
\label{hl2}
\end{equation}
If one expands the solution around $\lambda=0$, one finds that
$$h_\infty(\lambda)= 1 + \lambda + {1 \over 2} {m \over m-1} \lambda^2 + 
{1 \over 6} {m^2 (m+2) \over (m^2 -1)(m-1)} \lambda^3 + \dots $$
and the comparison with (\ref{hl}) gives for large $G$
$$ {\langle w^2 \rangle \over \langle w  \rangle^2} \to {m \over m-1}  \ \ \ 
\ ; \ \ \ \ {\langle w^3 \rangle \over \langle w  \rangle^3} \to {m^2 (m+2)  
\over (m^2 -1) (m-1)}   \ \ \  ; \ \ \ $$
which means that in principle the whole shape of $P(w)$ can be extracted from
(\ref{hl2}). 
In particular, one can  predict the power law of $P(w)$ at small $w$.
Trying to solve (\ref{hl2}) for  large  negative $\lambda$,   if
one writes
\begin{equation}
h_\infty(\lambda) - S(\infty) \simeq {1 \over | \lambda|^{\xi + 1} }
\label{hl3}
\end{equation}

one finds, as expected, that $S(\infty)$ is the fixed point of
(\ref{SG}). Eq. (\ref{hl3})
is equivalent to the asumption that $P(w) \sim w^{\xi}$ at small $w$,
where the exponent $\xi$ should satisfy
$$ 1 = S(\infty) m^{\xi + 2} \; . $$
This  completes the derivation of (\ref{beta}) which was already 
discussed in our previous work (Derrida {\it et al.}, 1999).

\subsection{Overlap between two trees}
Let us now show how (\ref{qab}) can be derived. Starting from recursion
(\ref{recursion}), one obtains by averaging over all the links relating 
generation $G$ to generation $G+1$ 

\begin{equation}\label{corr}
\langle w_\gamma^{(\alpha)} (G+1) w_\gamma^{(\beta)} (G+1) 
\rangle = {\langle k \rangle \over 4}  \langle w_\gamma^{(\alpha) }
(G) w_\gamma^{(\beta)} (G) \rangle + 
{\langle k(k-1) \rangle \over 4} 
\langle w_\gamma^{(\alpha)} (G) w_{\gamma'}^{(\beta)} (G) \rangle \; ,
\end{equation}
where $\gamma \neq \gamma'$ and the averages over $k$ are carried out with 
respect to (\ref{pkexact}). This gives 

\begin{equation}
\langle w_\gamma^{(\alpha)} (G+1) w_\gamma^{(\beta)} (G+1) 
\rangle =  {m \over 4} \langle w_\gamma^{(\alpha)} (G) w_\gamma^{(\beta)} 
(G) \rangle +
{1 \over 4} \left( m^2 - {m \over N(G+1)} \right) 
\langle w_\gamma^{(\alpha)} (G) w_{\gamma'}^{(\beta)} (G) \rangle \; .
\end{equation}
Using the fact that the sum  $\sum_{\gamma'} w_{\gamma'}^{(\beta)} (G)=1$,
so that $\langle w_\gamma^{(\alpha)} (G) \rangle = 1 / N(G)$ at all 
generations, one gets that

\begin{eqnarray}
&& \langle w_\gamma^{(\alpha)} (G+1) w_\gamma^{(\beta)} (G+1) \rangle = 
{m \over 4} \langle w_\gamma^{(\alpha)} (G) w_\gamma^{(\beta)} (G) 
\rangle \nonumber \\
&& \;\;\;\;\;\;\;\;\;\;\;\;\;\;\;\;\;\; + {1 \over 4} \left( m^2 - {m \over N(G+1)} \right) 
{{1 \over N(G)} - \langle 
w_\gamma^{(\alpha)} (G) w_{\gamma}^{(\beta)} (G) \rangle \over N(G)-1} \; .
\end{eqnarray}
Keeping only the dominant contributions for large $N$'s we arrive at

$$\langle w_\gamma^{(\alpha)} (G+1) w_\gamma^{(\beta)} (G+1) \rangle = 
{m \over 4} \langle w_\gamma^{(\alpha)} (G) w_\gamma^{(\beta)} (G) \rangle
+ {m^2 \over 4} {1 \over N(G)^2} \; . $$
Comparing this expression with (\ref{corr}), one sees that for large $N$, one 
could have  simply  neglected the correlations between the weights of 
different individuals, (i.e. directly replaced $ \langle w_\gamma^{(\alpha)} 
(G) w_{\gamma'}^{(\beta)} (G) \rangle  
$ by $ \langle w_\gamma^{(\alpha)} (G) \rangle  \ \langle  w_{\gamma'}^{(\beta)}
(G) \rangle$) and used the Poisson distribution instead of (\ref{pkexact})).
The previous recursion can be integrated 

\begin{equation}
\langle w_\gamma^{(\alpha)} (G) w_\gamma^{(\beta)} (G) \rangle = 
\Big[ \langle w_\gamma^{(\alpha)} (0) w_\gamma^{(\beta)} (0) \rangle  + 
{1 \over N^2} {m \over m-1}  (m^G-1) \Big] \left(m \over 4 \right)^G \; ,
\end{equation}
and using the fact that
$\langle w_\gamma^{(\alpha)} (G) w_\gamma^{(\beta)} (G) \rangle $  is  equal 
to $Y(G) / N(G)$ when $\alpha \neq \beta$ and  to $X(G)/ N(G)$ when $\alpha =
\beta$, one finds (with  $X(0)=1$ and $Y(0)=0$)

$$ {\langle Y(G) \rangle \over \langle X(G) \rangle } = {(m^{G} -1) m^{-G_c} 
\over 1 +  (m^{G} -1) m^{-G_c} } $$
where $G_c$ is given by (\ref{ngen}). For large $N$, that is for large $G_c$
this reduces to (\ref{qab}) in the whole range where the expression departs 
from $0$ or $1$, that is for $G$ of order $G_c$.
Finally, one can check that with the value of $G_c$ given by (\ref{ngen}), 
$N(G)$ is always large, as long as $N$ is large, so that our assumption that 
all the $N$'s were large was legitimate.

\vspace{0.5cm}
\noindent
ACKNOWLEDGEMENTS. The authors acknowledge discussions with Jordi Bascompte, 
Ugo Bastolla and Julio Rozas. SCM thanks the Alexander von Humboldt Foundation 
for support.

\begin{figure}
\centerline{\psfig{file=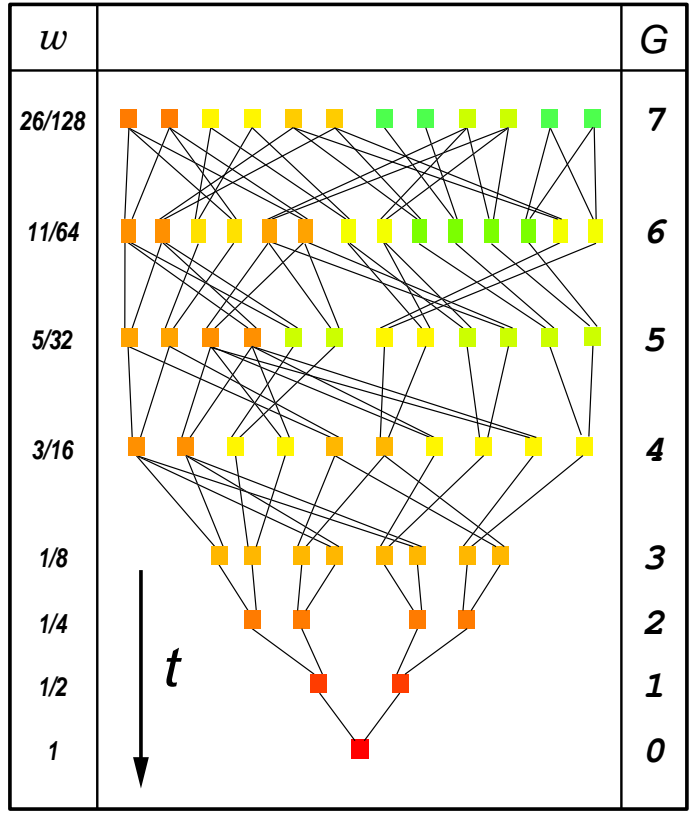,width=9cm,angle=0}}
\vspace{.5cm}
\caption{Coalescence of branches in a genealogical tree. We display the
reconstructed ancestry of a present individual in a small population of 
constant size $N=14$. 
Numbers on the left side stand for the {\it weight} $w$
of the leftmost individual at each generation. The grey scale changes 
from light grey (small $w$) to dark grey (large $w$) proportionally 
to the logarithm of the weight. The exact values are calculated 
according to Eq. (2). The weight is a measure proportional to the 
number of times that an ancestor appears in a tree, or, equivalently, 
to the number of branches which have coalesced up to that point. }
\end{figure}

\begin{figure}
\centerline{\psfig{file=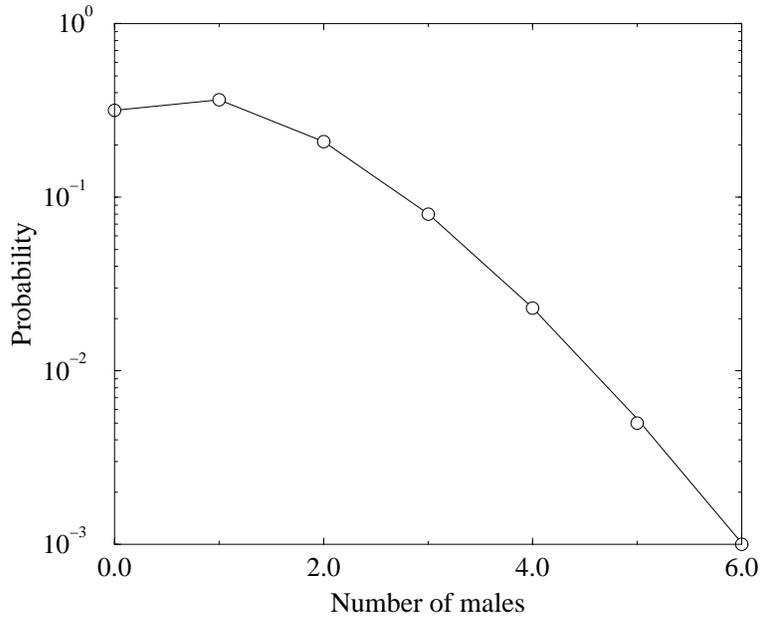,width=10cm,angle=270}}
\caption{Probability for an English couple to have $k$ marrying sons
during the period 1350-1986 (open circles). The solid line corresponds to 
a Poisson distribution of average $1.15$. (Data from Dewdney (1986)).}
\end{figure}

\begin{figure}
\centerline{\psfig{file=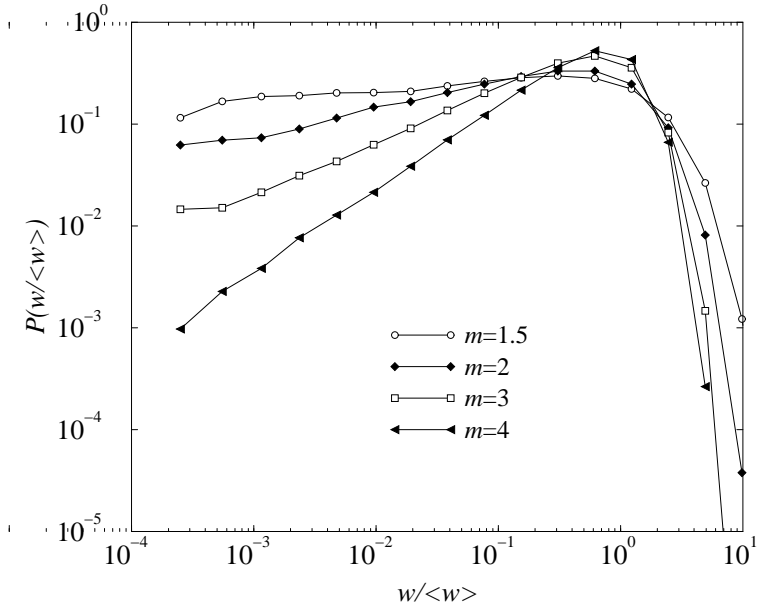,width=10cm,angle=270}}
\caption{ Stationary shape of the distribution
$P(w/\langle w \rangle) $ for different
values of $m$. We compare the constant population case ($m=2$) with 
shrinking ($m=1.5$), and expanding ($m=3, \; 4$) populations. Parameters
are $N=4096$, $G=20$, and averages over $10^3$ independent realizations 
have been performed. }                        
\end{figure}

\begin{figure}
\centerline{\psfig{file=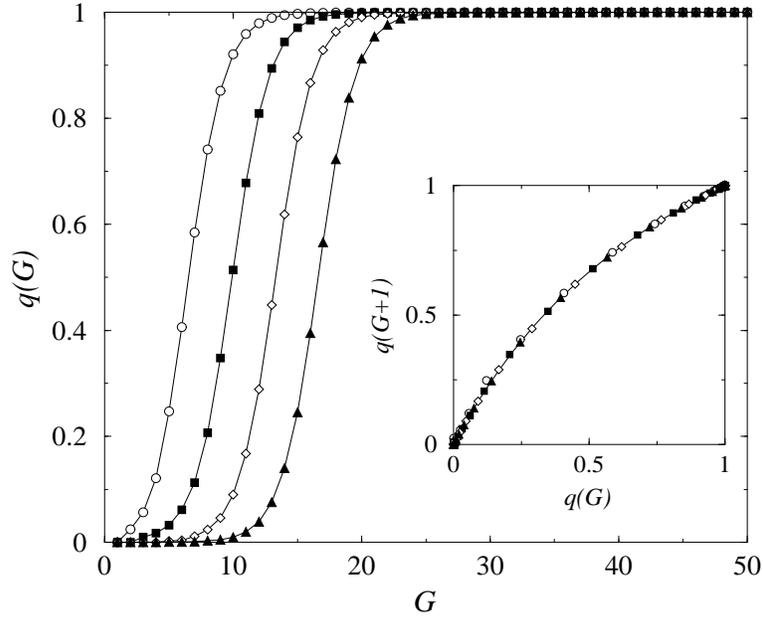,width=10cm,angle=270}}
\caption{The averaged overlap $q(G)$  as a function
of the number of generations $G$. The results of simulations for 
different sizes of the population  $N=100$ (open circles), $1000$ 
(solid squares), $10000$ (open diamonds), $100000$  (solid triangles) 
agree with this prediction, up to small  finite size corrections  only 
visible for $N=100$. The insert shows the results of simulations and 
the prediction (\ref{qevo}).}
\end{figure}

\begin{figure}
\centerline{\psfig{file=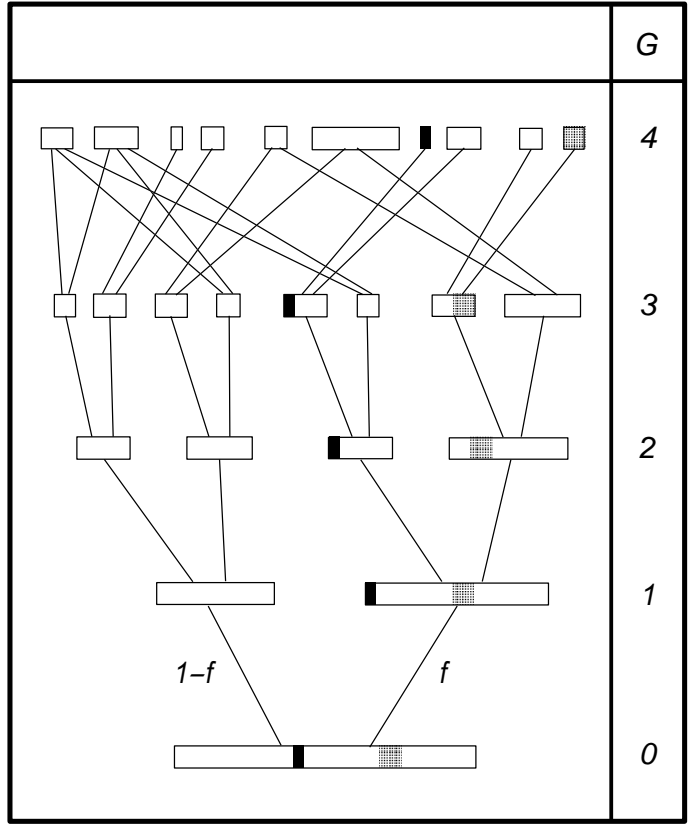,width=9cm,angle=0}}
\vspace{.5cm}
\caption{Representation of the first 5 generations of the tree in Fig. 1
with a random distribution of the weight of an individual between
its two parents. The fraction $f$ of the weight contributed by each
ancestor is randomly chosen from a distribution with average value 
$\langle f \rangle = 1/2$.}
\end{figure}

\begin{figure}
\centerline{\psfig{file=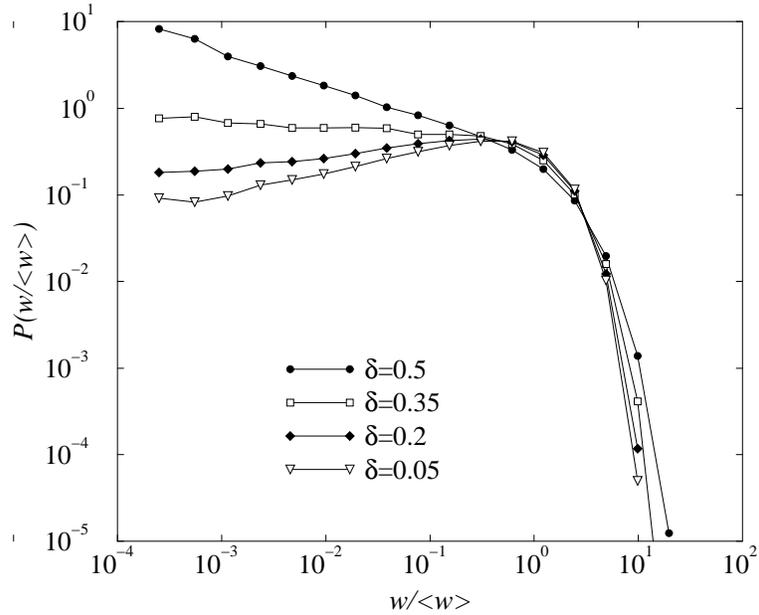,width=10cm,angle=270}}
\caption{Stationary distribution of weights  $P(w / \langle w \rangle )$
versus $w / \langle w \rangle$ for several choices of $\delta$.
The fixed population size is $N=4096$, and we have averaged over 
$10^3$ independent runs. Values of $\delta$ are as shown in the legend.}
\end{figure}

\begin{figure}
\centerline{\psfig{file=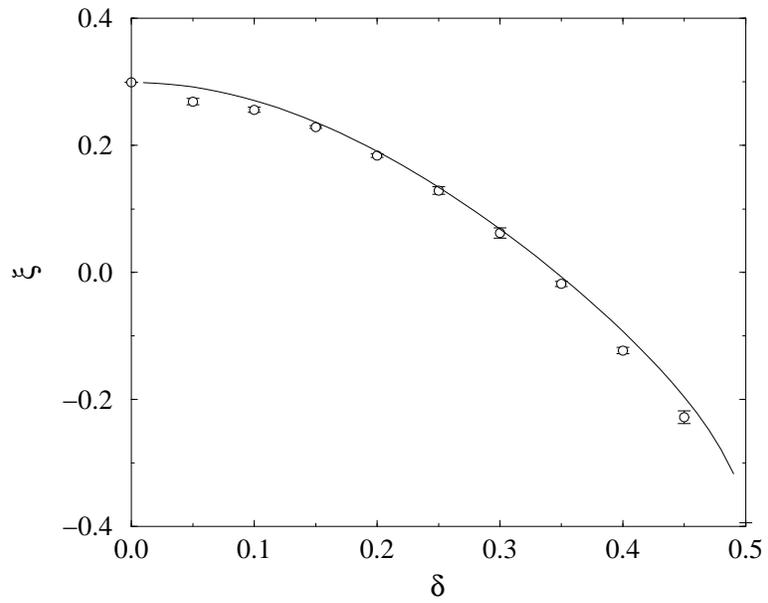,width=10cm,angle=270}}
\caption{Comparison between the predicted  values of the exponent
$\xi$ (solid line) given by (\ref{bd}) and the results of the simulations 
for different values of $\delta$ (circles). Parameters as in Fig. 6. 
For a value of $\delta \simeq 0.35$, the exponent $\xi$ changes sign.
This point is important since the typical contribution of a randomly 
chosen ancestor changes suddenly in a finite amount.}
\end{figure}

\end{document}